\def\year{2020}\relax
\documentclass[letterpaper]{article} % DO NOT CHANGE THIS
\usepackage{aaai20}  % DO NOT CHANGE THIS
\usepackage{times}  % DO NOT CHANGE THIS
\usepackage{helvet} % DO NOT CHANGE THIS
\usepackage{courier}  % DO NOT CHANGE THIS
\usepackage[hyphens]{url}  % DO NOT CHANGE THIS
\usepackage{graphicx} % DO NOT CHANGE THIS
\usepackage{graphicx}  % DO NOT CHANGE THIS
\usepackage[caption=false]{subfig}
\usepackage{amsmath}
\usepackage{algorithm}
\usepackage{algpseudocode}
\usepackage{multirow}
\usepackage{verbatim}
\usepackage{filecontents}
\usepackage{pgfplots}
\usepackage{pgfplotstable}
\usepackage{footnote}
\title{An Attention-Based Speaker Naming Method for Online Adaptation\\ in Non-Fixed Scenarios}
\author{Jungwoo Pyo\textsuperscript{\rm 1}, Joohyun Lee\textsuperscript{\rm 1}, Youngjune Park\textsuperscript{\rm 1}, \\ \Large \textbf{Tien-Cuong Bui\textsuperscript{\rm 1}, Sang Kyun Cha\textsuperscript{\rm 1}}\\ % All authors must be in the same font size and format. Use \Large and \textbf to achieve this result when breaking a line
\textsuperscript{\rm 1}Seoul National University, Seoul, Korea\\ %If you have multiple authors and multiple affiliations
\{wjddn1801, wngusdlekd, dudwns930, cuongbt91, chask\}@snu.ac.kr % email address must be in roman text type, not monospace or sans serif
}
\begin{document}

\maketitle
\begin{abstract}
A speaker naming task, which finds and identifies the active speaker in a certain movie or drama scene, is crucial for dealing with high-level video analysis applications such as automatic subtitle labeling and video summarization. Modern approaches have usually exploited biometric features with a gradient-based method instead of rule-based algorithms. In a certain situation, however, a naive gradient-based method does not work efficiently. For example, when new characters are added to the target identification list, the neural network needs to be frequently retrained to identify new people and it causes delays in model preparation. In this paper, we present an attention-based method which reduces the model setup time by updating the newly added data via online adaptation without a gradient update process. We comparatively analyzed with three evaluation metrics(accuracy, memory usage, setup time) of the attention-based method and existing gradient-based methods under various controlled settings of speaker naming. Also, we applied existing speaker naming models and the attention-based model to real video to prove that our approach shows comparable accuracy to the existing state-of-the-art models and even higher accuracy in some cases.

\end{abstract}

\section{Introduction}

\begin{figure}[ht]
    \centering
     \subfloat[]{%
        \includegraphics[width=\linewidth]{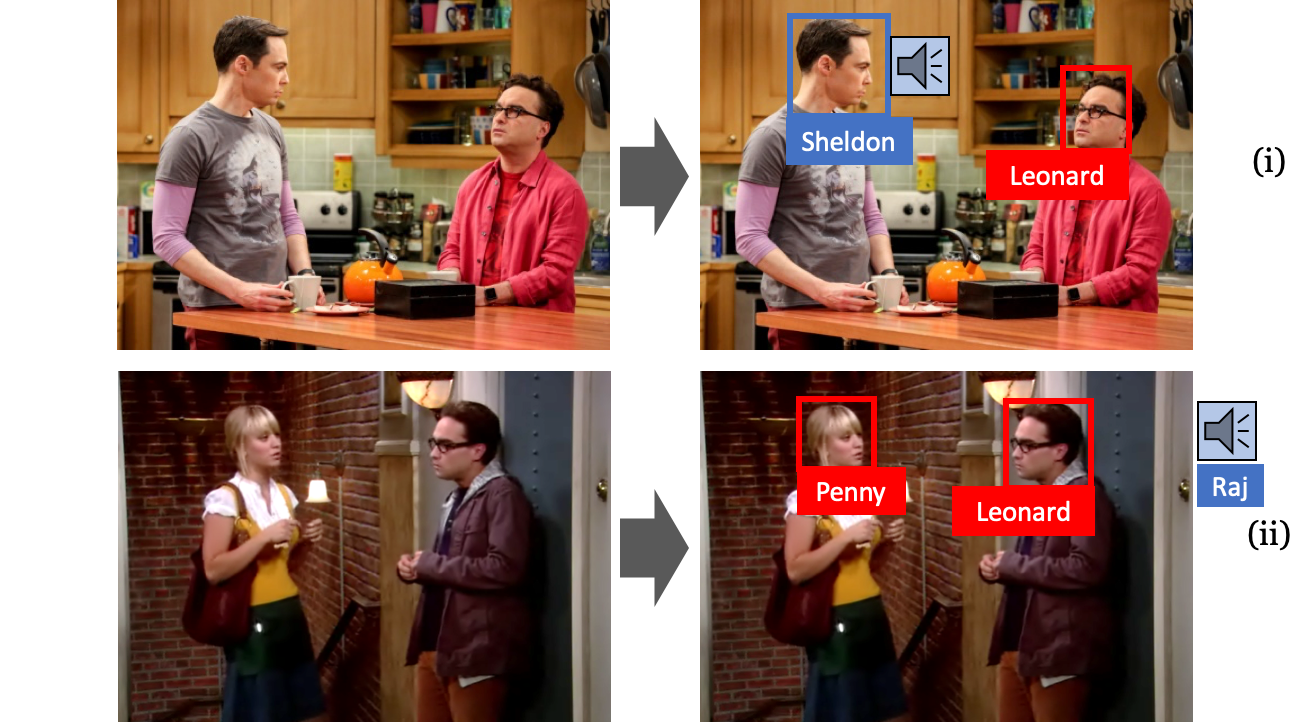}
    }
     \hfill
     \subfloat[]{%
        \includegraphics[width=\linewidth]{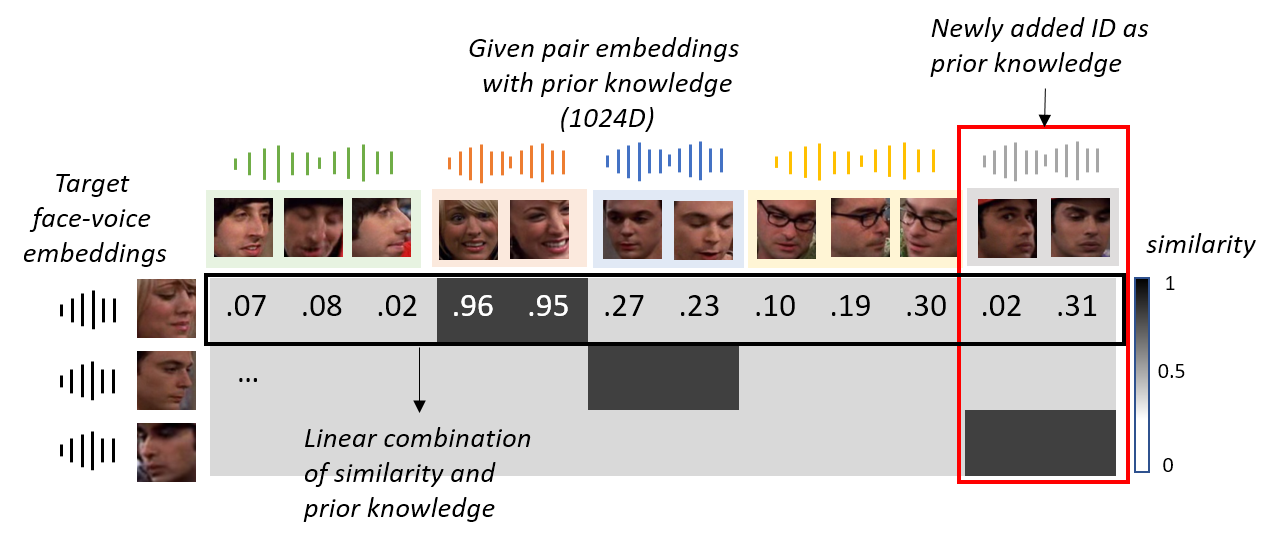}
    }
     \hfill
    \caption{(a) Speaker naming task contains two situations: i) finding matched face with corresponding voice if speaker appears in the scene, ii) picking out all distractors if speaker is out of the scene. (b) Visualization of predicting ID of face-voice pair with few-shot learning based attention module. The predicted ID of the target can be inferred to linear combination of cosine similarity between every prior knowledge embedding and target embedding. It also shows the simplicity of dealing with data of newly added ID inserting into the attention module.}
    \label{fig:intro}
\end{figure}

% biometric recognition, 그 중에서도 speaker naming이 중요한 문제인 이유
Biometric recognition plays an important role in advanced authentication systems. It identifies individuals based on physical or behavioral characteristics. The speaker naming task, which is to identify visible speaking characters in multimedia videos, consists of multiple types of biometric recognition. Most of the speaker naming methods distinguish an active speaker based on biometric features like face image or voice. The necessity of this task is proven to be essential for high-level video analysis problems such as summarization \cite{takenaka2012drive}, interaction analysis \cite{liu2008naming}, and semantic indexing \cite{zhang2013attribute}. In particular, identification of active speaking characters with automatic subtitling can help deaf audiences to enjoy the videos more without any difficulties in understanding the context.

%기존의 speaker naming method들이 어떤 부분에 초점을 맞추었고, 어떤 방법을 이용하여 문제를 해결하였는지
Most of existing speaker naming models mainly focus on boosting up the accuracy for finding active speaker among fixed origin character list. Gradient-based methods are considered one of the right solutions to get higher accuracy, and these methods have been proposed in various ways using multiple modalities for speaker naming. \cite{hu2015deep} proposes a deep multimodal model based on a CNN architecture to extract the facial and acoustic features from videos, then combine them through a fusion function. Correspondingly, a multimodal Long Short-Term Memory(LSTM) architecture\cite{ren2016look} merges visual and auditory modalities from the beginning of each input sequence. \cite{Bredin:2016:ISD:2964284.2967202} improves the performance of talking-face detection by capturing the lip motion. 

%변화하는 환경에 대해서 잘 대처하는 것도 중요한 고려사항임을 언급하고, 해당 method가 고정된 환경이 아닌 변화하는 환경에도 잘 대처할 수 있는지도 중요한 고려사항이다. Speaker naming task는 대부분의 multimedia에 적용 가능한 task이기 때문에 model이 특정 dataset에 dependent하면, 각기 다른 dataset마다 별개의 model을 이용하여야 하는 번거로움이 있는데, 이는 매우 비효율적이다. 
However, it is not always possible to occur expected or fixed situation in a video. In the real world, there are several uncertain situations which cause the difficulties of identifying active speaker such as appearing new characters or misinterpretation due to lack of labeled training data. Most of the gradient-based identification approaches cannot immediately adapt to a change in predicted character list or update the newly added data to model. Traditionally, these methods have to predefine a set of targeted characters before the training period. In particular, an existing model has to be retrained with a new set of targeted characters, which consist of both origin classes and the new ones, from scratch. It causes much time-consuming to rebuild the model. Transfer learning \cite{yosinski2014transferable} and domain adaptation\cite{Ganin:2015:UDA:3045118.3045244} are proved to be efficient for faster adaptation of neural network through initialization based on origin data. Nonetheless, these methods still require considerable time in the training phase adapting the newly added data to original model, which takes considerable time.
Besides, it is hard to contain a sufficient amount of labeled data in real-world datasets since labeling task is costly. The availability of labeled data poses a major practical issue for many gradient-based models.

% attention method를 소개하며,  online update가 가능하다는 점을 이용해 변화하는 상황에 적용하기 적합한 방법임을 주장함
To overcome these problems, we apply an attention module with few-shot learning \cite{fei2006one} for making our identification model flexible to accommodate changes at run-time. The attention module, which is based on scaled dot-product attention structure \cite{vaswani2017attention}, represents the similarities between prior knowledge embeddings and extracted features from the target video. The prior knowledge embeddings are the given data, which consist of facial and vocal embeddings of the predicted classes from the training dataset.
% The model aims at highlighting the similarity of these embedding matrices to maximize the prediction accuracy.
Next, few-shot learning is used for dealing with the scarcity of labeled data and imbalanced class distribution. Attention mechanism and few-shot learning are effectively combined in our model since they are both linear and straightforward. The essential component of the few-shot learning method derives feature embeddings based on a distance function. The attention mechanism consists of a linear combination with scaling and a softmax operation among these feature embeddings as shown in Figure \ref{fig:intro}(b). This combination makes the model consider every single embedding of prior knowledge carefully. Therefore, our method works well under the conditions even with a small amount of data or a highly imbalanced class distribution. More importantly, our model only utilizes the pretrained neural networks to extract embeddings. It means our model does not comprise a backpropagation process unlike other gradient-based models. Namely, the setup time is significantly decreased by updating the new information on the attention module in run-time.

%attention module의 단점도 서술할것(memory)
%추가적으로, 본 논문에서는, 다양한 조건을 가지고 있는 상황들을 설정하고 각 상황들에 적합한 speaker naming method들이 무엇인지를 3가지의 metric를 기반으로 비교.분석하는 실험을 수행하였다.
%각 상황에 따른 case study를 수행한다는 내용을 집어넣을 것
However, our proposed method is not always an optimal solution for all situations. In situations where character changes are not frequent or there are many IDs to identify, a deep-learning approach that guarantees robust performance may be more suitable even if the model setup takes a long time. Consequently, we compared attention-based method with gradient-based methods under various conditions of speaker naming by adjusting two variables: the number of target IDs to be identified, and the number of shots per each character. Furthermore, we compared our proposed model with existing speaker naming models on real video.

Our contributions are summarized as follows: 
\begin{itemize}
    \item We proposed a non-gradient-based method using attention module with few-shot learning, which can efficiently deal with the scarcity of labeled data as well as imbalanced class distribution.
    \item Our model significantly reduces the setup time of the model by removing the gradient descent process and updating the new data to model online.
    \item Under various environments adjusting both the number of target IDs to be identified and the number of shots per each character, we conducted comparative analyses with real-world dataset between our proposed method and existing gradient-based methods through three metrics: accuracy, memory usage, and setup time.
    \item Our model shows comparable accuracy to the state-of-the-art speaker naming models on real video.
\end{itemize}

\section{Related Work} \label{related_work}
\begin{figure*}[ht!]
\begin{center}
  \includegraphics[width=.9\textwidth]{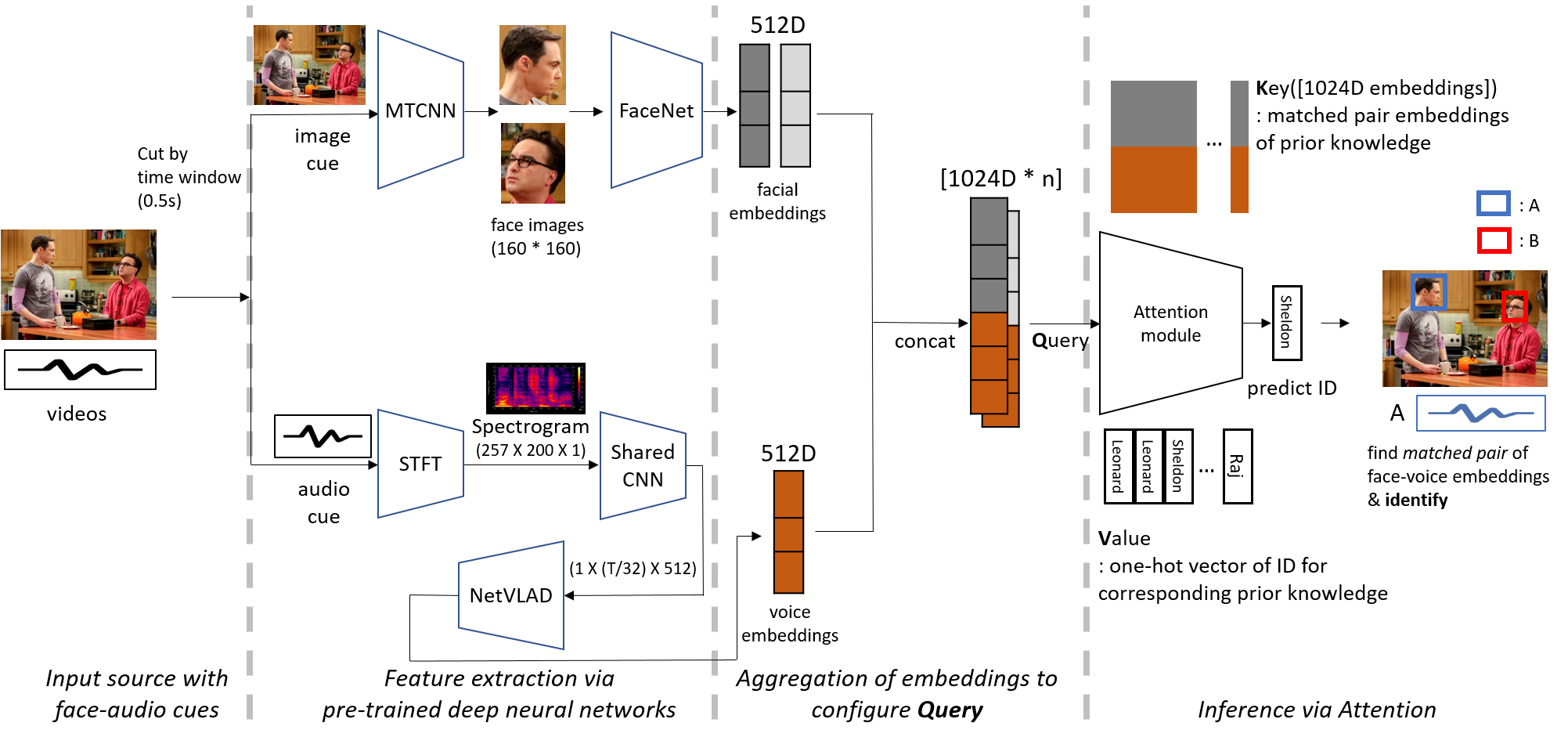}
  \caption{Overall architecture of attention-based model for speaker naming.}
  \label{fig:overall}
\end{center}
\end{figure*}

\subsection{Speaker Naming}
%speaker naming task 정의, 논문 list, 각 논문별 해결 방식
% 대부분의 speaker naming task들은 matched-pair non-matched-pair 방식을 이용하고, final layer 부분을 fine-tuning하는 방식으로 진행(우리는 fine-tuning을 하지 않는다는 것을 어필)
Speaker naming is a task to identify the speaker in a video source. Recent studies about automatic speaker naming used deep neural networks to get each speaker's name from the multimodal face-voice sources. In \cite{hu2015deep}, they proposed a convolutional neural network (CNN) based multimodal framework that automatically learns from the combined face and acoustic features. They trained the SVM classifier to reject all non-matching face-voice pairs and get identification results. Likewise, \cite{ren2016look} improved the accuracy by changing the CNN based model to Long Short-Term Memory (LSTM) based model. This change gave robust identification results for face distortion. \cite{LIU2019557} used attention architecture to accommodate the face variations.

\subsection{Feature Extractors for Face and Audio Cues}
The primary purpose of feature extractors is to express a particular type of data to distilled numerical embeddings, which has lower dimension than original data. Several feature extractors have been studied in each field according to various types of data. Most feature extractor operates by setting appropriate loss function and distance metric, then optimizing them.

\subsubsection{Face}
%Face extractor: facenet, arcface non-linearity
Various types of loss functions have been tried to use as facial feature extraction. \cite{sun2014deep,wen2016discriminative} used cross-entropy loss to minimize euclidean distance. FaceNet\cite{schroff2015facenet} introduced triplet loss based on Euclidean distance to train the face feature extractor. Also, FaceNet utilized MTCNN\cite{zhang2016joint} to extract the aligned cropped face images from raw image dataset. SphereFace\cite{liu2017sphereface}, CosFace\cite{wang2018cosface}, ArcFace\cite{deng2019arcface} used angular loss to minimize the cosine similarity.

\subsubsection{Audio}
%Voice extractor: MFCC, CNN, NetVLAD
The area of feature extraction of audio data has also been studied in various directions. There have been many useful methods, such as using MFCC \cite{muda2010voice} and using CNN \cite{hershey2017cnn}. Recently, \cite{xie2019utterance} suggested a new model using "thinResNet" trunk architecture and dictionary-based NetVLAD layer. This method successfully performed the speaker identification task on audio data with varying lengths and mixing of unrelated signals.

\subsection{Attention Mechanism}
% Attention은 원래 ~한 목적으로 제안되었다는 것을 먼저 설명
Attention mechanism is first proposed in \cite{bahdanau2014neural} for neural machine translation(NMT) field. The attention mechanism looks up all of the input elements(e.g., sequential input such as frames in video or words in a sentence) at every decoding time, calculates the attention map which is a matrix that reflects the relevance of present input and previous input elements.

% attention의 variant들에 대하여 설명
Attention map is a probability matrix that each target word is aligned to, or translated from source word. Each element of the attention map is computed as the softmax value, which means the similarity of source word and each target word.

% Attention의 variant들에 대한 소개
Some papers have brought attention mechanism to speaker naming task. In \cite{LIU2019557}, they proposed attention guided deep audio-face fusion approach to detect active speaker. They also used individual network model to convert from face and voice sources to each embedding as ours. Before fusing the face and voice embeddings, they applied the attention module only for face embeddings to consider the relationship between other face embeddings. However, our work applied attention mechanism to fusion of face-voice pair embeddings and focused on the relevance of target embeddings and prior knowledge embeddings.  

\section{Methodology} \label{methodology}

%% 문제를 formulation, 기존의 방식은 related work에 있으므로 쓰지 말 것
Speaker naming contains all processes from detecting faces, recognizing voice, and matching these embeddings to identifying the current speaker. As shown in Figure \ref{fig:intro}(a), we regard the speaker naming problem in two cases. The first case is to find out the pair embeddings that both face and voice embedding are identified as same ID(so-called "matched-pair"). The second one is to pick out the pair embeddings where ID of face and voice do not match(so-called "non-matched-pair"). We propose a non-gradient based method using attention networks with few-shot learning to solve this problem. In this section, we formulate our problem precisely and elaborate on our proposed model.  

\begin{figure}[ht!]

  \includegraphics[width=.9\linewidth]{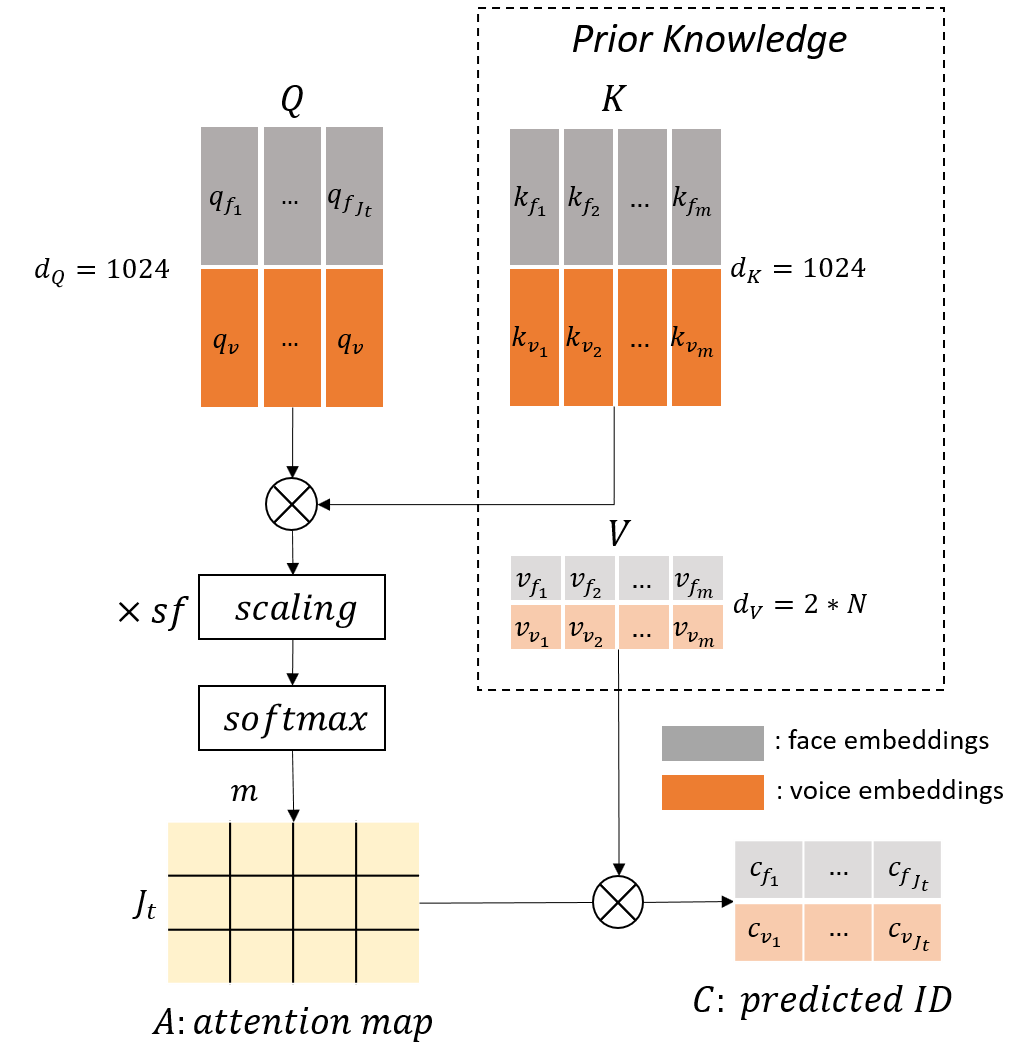}
  \caption{Mechanism of attention module with few-shot learning.}
  \label{fig:attention_mechanism}

\end{figure}
\subsection{Problem Formulation} \label{methodology:problem_formulation}
%formulation
We formulate our problem as follows. Let $t$ be the index of time window, $I=\{i_1, i_2, ..., i_N\}$ denotes the ID of characters. $J_t$ is the number of faces which are captured in $t$. $f_j^{t}$ is the $j$-th number of face embeddings cropped in time window $t$. Likewise, $v_{t}$ represents the voice embedding in time window $t$. Then the maximum probability of facial embedding whose ID is $i_k$ in time window $t$ is as follows.
\begin{equation} \label{eqn: f_prob}
    F_{prob}(i_k, t) = \underset{1 \leq j \leq J}{\rm{max}} p(i_k | f_j^{t})
\end{equation}
By multiplying $F_{prob}(i_k, t)$ and the probability that predicted ID of voice embedding in $t$ is $i_k$, we can infer the ID of speaker in $t$ as below.
\begin{equation} \label{eqn: spk_id}
    Spk_{ID}(t) = \underset{i_k \in I}{\mathrm{argmax}}(F_{prob}(i_k, t) \cdot p(i_k | v^{t}))
\end{equation}
Based on Equation \eqref{eqn: spk_id}, we calculate the accuracy of the speaker naming model if it correctly estimates the ID of matched-pair, or picks out the non-matched-pair in the time window $t$. After all, we aggregate $Spk_{ID}(t)$ over all time windows to get the total accuracy of the target video.

\subsection{Attention-Based Method for Speaker Naming} \label{methodology:proposed_model}
% K,V같은 것들이 너무 많아서 가독성이 떨어짐, K - V가 뺴긴지 뭔지 모호함
% 우리 네트워크가 풀고자 하는 문제 및 간략한 방법
% 우리는 위 문제를 어떤 식으로 풀어갈 건지 간단 요약
%speaker naming이 기본적으로 matched pair를 찾는 task라는 것을 설명
The speaker naming problem consists of two parts: finding out matched face-voice pairs to predict current speaker, and picking out the non-matched-pairs. Our approach to solve the problem is as follows. First, we capture the face images and voice chunks by every fixed size of the time window. Then, we convert face images and voice chunks to extracted embeddings with pre-trained face and audio feature extractor. We concatenate both face and voice embeddings to make candidates of pair embeddings by each frame. Then, we calculate the attention map with this concatenated embedding. Attention map applies scaling and a softmax function to cosine similarity matrix among all of the characters' prior knowledge embeddings and extracted target embeddings. We predict the IDs of target embeddings based on the attention mechanism. Then, the proposed method aggregates the prediction result by each time window, and it determines the active speaker in the scene. Finally, we measure the prediction accuracy of the model by aggregating every result of all time windows. We describe our method's overall architecture and flow in Figure \ref{fig:overall}.

\textbf{Feature Extraction. }
% face & audio
For generating embeddings which contain the features of facial appearance and voice, we use pre-trained feature extractors, which convert raw input sources to numerical vectors with reduced dimensions.
Our network uses FaceNet as facial feature extractor, NetVLAD as voice feature extractor. 
The weights of these extractors are fixed while updating the attention module or end-to-end inferencing phase. 

% Attention 모듈을 사용하는 이유를 논리적으로 서술
%% Q,K,V를 기반으로 한 speaker naming에서의 구체적인 설명
\textbf{Attention Module with Few-Shot Learning.}
Our attention module with few-shot learning consists of multiple components. Let $Q$ denotes the query matrix, which is the extracted face-voice pair embeddings from target video. $Q$ contains several matched-pairs and non-matched-pairs, which we will predict within a certain time window. Thus, $Q$ is a variable for time window $t$. $K$, $V$ belong to prior knowledge for our network. $K$ denotes the matrix of multiple face-voice pair embeddings extracted from the training data. $V$ is a one-hot vector matrix of IDs corresponding to $K$. These $KV$ set work as proofs for our decision whether the pair embeddings are matched-pair or not, and classifying the pair's ID. 

Detailed process of attention mechanism is shown in Figure \ref{fig:attention_mechanism}. The intuitive role of attention module is to consider the correlations between every pair of $Q$ and $K$. In our case, computing attention map and context vectors in attention module correspond to computing similarity and matrix of predicted IDs, respectively. As a distance metric, we use cosine distance because embeddings in $Q$ and $K$ are unit vectors, we can get cosine similarity using inner product. Before performing matrix multiplication of $Q$ and $K$, we use transposed matrix $Q^T$ to match the dimension.
Our method performs a few additional operations after doing matrix multiplication with $Q^T$ and $K$. First, multiply scale factor $sf$ to $Q^{T}K$. Then, apply the softmax function to all elements. Our network is set to the value of $sf$ as $\sqrt{d_K}$, where $d_K$ is the dimension of $K$, set to 1024. The reason for multiplying $sf$ is that after performing multiplication between unit vectors to calculate cosine similarity, the value becomes so small that it interferes with subsequent softmax operation. If the input parameter's scale of softmax function is too small or large, it cannot express the appropriate probability distribution. Our setting can arrange the scale of value at the proper level to perform softmax.   
Based on the above explanation, the attention map is mathematically written as:
\begin{equation} \label{eqn: attention_map}
    A = \mathrm{softmax}(sf(Q^TK))
\end{equation}
The context vectors which represents the prediction of IDs to $Q$ is written as:
\begin{equation} \label{eqn: prediction}
    C = VA^T
\end{equation}
% confidence score 구하는 방법 설명
$C$ represents the probability of which face-voice pair in $Q$ is regarded as a particular ID. The probability of face and voice are separated in $C$. Our method uses confidence score vector $c_p$ as the criteria for the decision to distinguish whether the $p$-th embeddings in $Q$ is the active speaker or not.  We apply Hadamard product\cite{horn1990hadamard}, which multiplies the face part and voice part of $C$ element-wise, to consider both features of face and voice. From this operation, we get $1 \times N$ vector of confidence score where $N$ is the number of IDs. The maximum value and its index in  $c_p$ will be regarded as a confidence score and $Spk_{ID}(t)$, respectively.
We elaborate on the procedure of overall flow in Algorithm \ref{alg: SN}. 

%Algorithm
\begin{algorithm}[htb!]
\begin{algorithmic}[]
\caption{End-to-End Speaker Naming Prediction} \label{alg: SN}
\State \textbf{Let} $Q$: Query, $K$: Key, $V$: Value, $A$: Attention map
\State $I \gets {\{i_1, i_2, ..., i_N\}}$ : a set of characters' IDs
\State Cut video by 0.5s interval of time window 
\For {each time window $t \gets 1$ to $T$}
    \State $Rep_t$: representative frame in $t$
    \State $\{f_1, ..., f_{J_t}\}$: cropped $J_t$ faces from $Rep_t$
    \State $\{q_{f_1}, ... , q_{f_{J_t}}\}$: facial embeddings from $\{f_1, ..., f_{J_t}\}$
    \State $q_v$: voice embedding extracted from audio in $t$
    \State $Q \gets \begin{pmatrix} q_{f_1} & q_{f_2} & ... & q_{f_{J_t}} \\ q_v & q_v & ... & q_v \end{pmatrix}$
    \State $A \gets \mathrm{softmax}(sf(Q^TK))$
    \State $C \gets VA^T$
    %\State \textbf{return} $(c_{f} \cdot c_{v})$
    \State $max\_conf \gets 0$
    \For{$p \gets 1$ to $J$}
        \State $c_p \gets c_{f_p} \odot c_{v_p}$\Comment{Hadamard product}
        \State $max\_conf \gets$ max($max\_conf, max(c_p)$)
        \If{$max\_conf$ == $max(c_p)$}
            \State $Spk_{ID}(t) \gets \underset{i \in I}{\mathrm{argmax}}(c_p)$
        \EndIf
    \EndFor
\EndFor
\end{algorithmic}
\end{algorithm}

\section{Experiments} \label{experiments}

\subsection{Dataset Overview}
In experiments, we used two public datasets: the utterance videos of celebrities(\textit{VoxCeleb2}\cite{chung2018voxceleb2}) and a TV show(\textit{The Big Bang Theory(BBT)}). For the experiment in incremental settings, we randomly chose 500 people from \textit{VoxCeleb2}, which contain more than 10 videos. Then, we split the train set and valid set by 5:2 per each ID. For \textit{BBT}, we selected 5 episodes (\textit{S01E02, S01E03, S01E04, S01E05, S01E06}). Each episode consists of the whole video, face images with various poses and illumination, and aggregated voice file without silence. 

\subsection{Data Preprocessing}
%MTCNN, 데이터 추출, voxceleb의 경우 상위 500개 뽑아서 실험 Face, Audio feature extractor 
We used FaceNet and NetVLAD, which are the same extractors used in our model, for extracting train and test embeddings from the raw datasets. First of all, \textit{BBT} dataset consists of multiple cropped face images and merged voice files per ID, by each episode. Cropped face images were resized to $160 \times 160$ for fitting the image size with the input of FaceNet model. After that, we used pre-trained FaceNet to convert resized face images into 512-dimension embeddings. Similarly, we converted our audio file into 512-dimension voice embeddings. The window size of each audio chunk is 2s, cut with 0.1s stride.
We did additional preprocessing in order to get cropped face images and voice chunks when preprocessing \textit{VoxCeleb2} because it only consists of video files. First of all, we cut the videos every 30 frames per second. Then, MTCNN\cite{zhang2016joint} cropped face images for all frames. The captured images which are not actual face image were removed. About voice file, we applied the same setting of what \textit{BBT} was preprocessed.

\usetikzlibrary{pgfplots.groupplots}
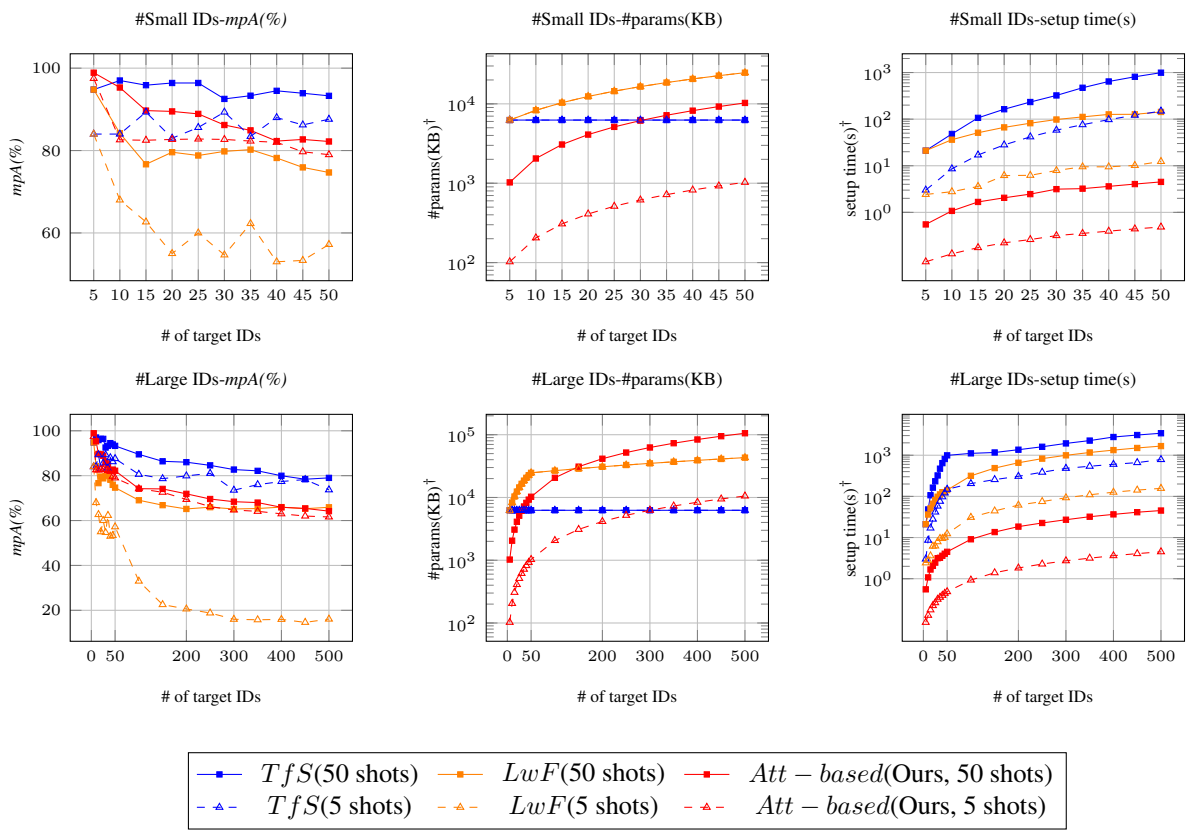
\begin{figure*}[ht!]
\centering
\pgfplotsset{
title style={font=\scriptsize},
tick label style={font={\fontsize{6.5pt}{6.5pt}\selectfont}},
label style={font={\fontsize{6.5pt}{6.5pt}\selectfont}},
}
\begin{tikzpicture}
    \begin{groupplot}[group style = {columns=3, rows=2, horizontal sep=0.1\linewidth, vertical sep=0.1\linewidth}, width=0.3\linewidth]

    \nextgroupplot[title = {\#Small IDs-\textit{mpA(\%)}},
            xlabel={\# of target IDs},
            xtick={5,10,15,20,25,30,35,40,45,50},
            ylabel={\textit{mpA(\%)}},
            ylabel style={yshift=-15pt},
            grid=major,
            skip coords between index={10}{19},
            legend style = {column sep=5pt, legend columns = 3, legend to name = grouplegend}
            ]
              \addplot+[color=blue, mark size=1pt, mark=square*, mark options={blue}] table[x=num_of_id, y expr=\thisrowno{1}*100]{id_acc.dat}; 
              \addlegendentry{$TfS$(50 shots)}
              \addplot+[color=orange, mark size=1pt, mark=square*, mark options={orange}] table[x=num_of_id, y expr=\thisrowno{2}*100]{id_acc.dat}; 
              \addlegendentry{$LwF$(50 shots)}
              \addplot+[color=red, mark size=1pt, mark=square*, mark options={red}] table[x=num_of_id, y expr=\thisrowno{3}*100]{id_acc.dat}; 
              \addlegendentry{$Att-based$(Ours, 50 shots)}
              
              \addplot+[color=blue, mark size=1.5pt, mark=triangle, mark options={blue,solid}, dashed] table[x=num_of_id, y expr=\thisrowno{1}*100]{id_acc_smallshot.dat}; 
              \addlegendentry{$TfS$(5 shots)}
              \addplot+[color=orange, mark size=1.5pt, mark=triangle, mark options={orange,solid}, dashed] table[x=num_of_id, y expr=\thisrowno{2}*100]{id_acc_smallshot.dat}; 
              \addlegendentry{$LwF$(5 shots)}
              \addplot+[color=red, mark size=1.5pt, mark=triangle, mark options={red,solid}, dashed] table[x=num_of_id, y expr=\thisrowno{3}*100]{id_acc_smallshot.dat}; 
              \addlegendentry{$Att-based$(Ours, 5 shots)}
    \nextgroupplot[title = {\#Small IDs-\#params(KB)},
            xlabel={\# of target IDs},
            xtick={5,10,15,20,25,30,35,40,45,50},
            ylabel={\#params(KB)\textsuperscript{\textdagger}},
            ylabel style={yshift=-15pt},
            ymode=log,
            grid=major,
            skip coords between index={10}{19}
            ]
            \addplot+[color=blue, mark size=1pt, mark=square*, mark options={blue}] table[x=num_of_id, y=tfs]{id_params.dat};
              \addplot+[color=orange, mark size=1pt, mark=square*, mark options={orange}] table[x=num_of_id, y=lwf]{id_params.dat};
              \addplot+[color=red, mark size=1pt, mark=square*, mark options={red}] table[x=num_of_id, y=att]{id_params.dat};
              \addplot+[color=blue, mark size=1.5pt, mark=triangle, mark options={blue,solid}, dashed] table[x=num_of_id, y=tfs]{id_params_smallshot.dat};
              \addplot+[color=orange, mark size=1.5pt, mark=triangle, mark options={orange,solid}, dashed] table[x=num_of_id, y=lwf]{id_params_smallshot.dat};
              \addplot+[color=red, mark size=1.5pt, mark=triangle, mark options={red,solid}, dashed] table[x=num_of_id, y=att]{id_params_smallshot.dat};
                   
    \nextgroupplot[title = {\#Small IDs-setup time(s)},
            xlabel={\# of target IDs},
            xtick={5,10,15,20,25,30,35,40,45,50},
            ylabel={setup time(s)\textsuperscript{\textdagger}},
            ylabel style={yshift=-15pt},
            ymode=log,
            ytick={1,10,100,1000,10000},
            grid=major,
            skip coords between index={10}{19},
            ]
              \addplot+[color=blue, mark size=1pt, mark=square*, mark options={blue}] table[x=num_of_id, y=tfs]{id_time.dat}; 
              \addplot+[color=orange, mark size=1pt, mark=square*, mark options={orange}] table[x=num_of_id, y=lwf]{id_time.dat}; 
              \addplot+[color=red, mark size=1pt, mark=square*, mark options={red}] table[x=num_of_id, y=att]{id_time.dat}; 
              \addplot+[color=blue, mark size=1.5pt, mark=triangle, mark options={blue,solid}, dashed] table[x=num_of_id, y=tfs]{id_time_smallshot.dat};
              \addplot+[color=orange, mark size=1.5pt, mark=triangle, mark options={orange,solid}, dashed] table[x=num_of_id, y=lwf]{id_time_smallshot.dat};
              \addplot+[color=red, mark size=1.5pt, mark=triangle, mark options={red,solid}, dashed] table[x=num_of_id, y=att]{id_time_smallshot.dat};

    \nextgroupplot[title = {\#Large IDs-\textit{mpA(\%)}},
            xlabel={\# of target IDs},
            xtick={0,50,200,300,400,500},
            ylabel={\textit{mpA(\%)}},
            ylabel style={yshift=-15pt},
            grid=major,
            ]
              \addplot+[color=blue, mark size=1pt, mark=square*, mark options={blue}] table[x=num_of_id, y expr=\thisrowno{1}*100]{id_acc.dat}; 
              \addplot+[color=orange, mark size=1pt, mark=square*, mark options={orange}] table[x=num_of_id, y expr=\thisrowno{2}*100]{id_acc.dat}; 
              \addplot+[color=red, mark size=1pt, mark=square*, mark options={red}] table[x=num_of_id, y expr=\thisrowno{3}*100]{id_acc.dat}; 
              
              \addplot+[color=blue, mark size=1.5pt, mark=triangle, mark options={blue,solid}, dashed] table[x=num_of_id, y expr=\thisrowno{1}*100]{id_acc_smallshot.dat}; 
              \addplot+[color=orange, mark size=1.5pt, mark=triangle, mark options={orange,solid}, dashed] table[x=num_of_id, y expr=\thisrowno{2}*100]{id_acc_smallshot.dat}; 
              \addplot+[color=red, mark size=1.5pt, mark=triangle, mark options={red,solid}, dashed] table[x=num_of_id, y expr=\thisrowno{3}*100]{id_acc_smallshot.dat}; 
              
            %   \addlegendimage{blue, mark size=1pt, mark=square*, mark options={blue}}
            %   \addlegendentry{$TfS$(500 IDs)}
            %   \addlegendimage{orange, mark size=1pt, mark=square*, mark options={orange}}
            %   \addlegendentry{$LwF$(500 IDs)}
            %   \addlegendimage{red, mark size=1pt, mark=square*, mark options={red}}
            %   \addlegendentry{$Att-based$(Ours, 500 IDs)}
            %   \addlegendimage{blue, dashed, mark size=2pt, mark=x, mark options={solid,blue}}
            %   \addlegendentry{$TfS$(50 IDs)}
            %   \addlegendimage{orange, dashed, mark size=2pt, mark=x, mark options={solid,orange}}
            %   \addlegendentry{$LwF$(50 IDs)}
            %   \addlegendimage{red, dashed, mark size=2pt, mark=x, mark options={solid,red}}
            %   \addlegendentry{$Att-based$(Ours, 50 IDs)}
    \nextgroupplot[title = {\#Large IDs-\#params(KB)},
            xlabel={\# of target IDs},
            xtick={0,50,200,300,400,500},
            ylabel={\#params(KB)\textsuperscript{\textdagger}},
            ylabel style={yshift=-15pt},
            ymode=log,
            grid=major,
            ]
              \addplot+[color=blue, mark size=1pt, mark=square*, mark options={blue}] table[x=num_of_id, y=tfs]{id_params.dat};
              \addplot+[color=orange, mark size=1pt, mark=square*, mark options={orange}] table[x=num_of_id, y=lwf]{id_params.dat};
              \addplot+[color=red, mark size=1pt, mark=square*, mark options={red}] table[x=num_of_id, y=att]{id_params.dat};
              \addplot+[color=blue, mark size=1.5pt, mark=triangle, mark options={blue,solid}, dashed] table[x=num_of_id, y=tfs]{id_params_smallshot.dat};
              \addplot+[color=orange, mark size=1.5pt, mark=triangle, mark options={orange,solid}, dashed] table[x=num_of_id, y=lwf]{id_params_smallshot.dat};
              \addplot+[color=red, mark size=1.5pt, mark=triangle, mark options={red,solid}, dashed] table[x=num_of_id, y=att]{id_params_smallshot.dat};
    \nextgroupplot[title = {\#Large IDs-setup time(s)},
            xlabel={\# of target IDs},
            xtick={0,50,200,300,400,500},
            ylabel={setup time(s)\textsuperscript{\textdagger}},
            ylabel style={yshift=-15pt},
            ymode=log,
            ytick={1,10,100,1000,10000},
            grid=major,
            ]
              \addplot+[color=blue, mark size=1pt, mark=square*, mark options={blue}] table[x=num_of_id, y=tfs]{id_time.dat};
              \addplot+[color=orange, mark size=1pt, mark=square*, mark options={orange}] table[x=num_of_id, y=lwf]{id_time.dat};
              \addplot+[color=red, mark size=1pt, mark=square*, mark options={red}] table[x=num_of_id, y=att]{id_time.dat};
              \addplot+[color=blue, mark size=1.5pt, mark=triangle, mark options={blue,solid}, dashed] table[x=num_of_id, y=tfs]{id_time_smallshot.dat};
              \addplot+[color=orange, mark size=1.5pt, mark=triangle, mark options={orange,solid}, dashed] table[x=num_of_id, y=lwf]{id_time_smallshot.dat};
              \addplot+[color=red, mark size=1.5pt, mark=triangle, mark options={red,solid}, dashed] table[x=num_of_id, y=att]{id_time_smallshot.dat};
    
    \end{groupplot}
    \node at ($(group c2r2) + (0,-3.5cm)$) {\ref{grouplegend}};
\end{tikzpicture}
\caption{Comparative Analyses between speaker naming methods under various settings. Three metrics(\textit{mpA}, the number of parameters, setup time) are measured for each situation where the number of target IDs and the number of shots per character are changed. \textsuperscript{\textdagger}The y-axis of the graph is in logarithmic scale.}
\label{fig:comp_analysis}
\end{figure*}

\makesavenoteenv{table}
\makesavenoteenv{tabular}
\begin{table*}[h] 
    \centering
    \resizebox{.9\textwidth}{!}{
    \begin{tabular}{|c|c|c|c|c|c|}
        \hline
           \# of target IDs & \# of shots per each ID & metric & \textit{Training from Scratch(TfS)} & \textit{Learning without Forgetting(LwF)} & \textit{Att-based}(Ours) \\
        \hline
        % \multirow{6}{*}{Small IDs(5-50 people)}& \multirow{3}{*}{Small shots(5 shots)} & \textit{mpA}(\%)       & \textbf{+++}\textsuperscript{\textasteriskcentered} & + & ++ \\
        %                         &           & \# of params(KB)       & ++(constant) & +(linear)
        %                         & \textbf{+++(quadratic)} \\
        %                         &           & Setup time(s) & + & ++ & \textbf{+++} \\\cline{2-6}
        %                         & \multirow{3}{*}{Large shots(50 shots)} & \textit{mpA}(\%)       & \textbf{+++} & + & ++ \\
        %                         &                           & \# of params(KB)       & \textbf{+++(constant)} & ++(linear) & \textbf{+++(quadratic)} \\
        %                         &                           & Setup time(s) & + & ++ & \textbf{+++} \\\cline{1-6}
        % \multirow{6}{*}{Large IDs(50-500 people)}& \multirow{3}{*}{Small shots(5 shots)} & \textit{mpA}(\%)       & \textbf{+++} & + & ++ \\
        %                         &           & \# of params(KB)       & \textbf{+++(constant)} & ++(linear) & \textbf{+++(quadratic)} \\
        %                         &           & Setup time(s) & + & ++ & \textbf{+++} \\\cline{2-6}
        %                         & \multirow{3}{*}{Large shots(50 shots)} & \textit{mpA}(\%)       & \textbf{+++} & ++ & + \\
        %                         &                           & \# of params(KB)       & \textbf{+++(constant)} & ++(linear) & +(quadratic) \\
        %                         &                           & Setup time(s) & + & ++ & \textbf{+++} \\
        \multirow{6}{*}{Small IDs(5-50 people)}& \multirow{3}{*}{Small shots(5 shots)} & \textit{mpA}(\%)       & \textbf{86.05} & 61.02 & 83.38 \\
                                &           & \# of params(KB)       & 6248.4(constant) & 15473.4(linear)
                                & \textbf{564.39(quadratic)} \\
                                &           & Setup time(s) & 60.02 & 7.06 & \textbf{0.29} \\\cline{2-6}
                                & \multirow{3}{*}{Large shots(50 shots)} & \textit{mpA}(\%)       & \textbf{94.80} & 80.29 & 88.06 \\
                                &                           & \# of params(KB)       & 6248.4(constant) & 15473.4(linear) & \textbf{5644.03(quadratic)} \\
                                &                           & Setup time(s) & 381.21 & 86.28 & \textbf{2.64} \\\cline{1-6}
        \multirow{6}{*}{Large IDs(50-500 people)}& \multirow{3}{*}{Small shots(5 shots)} & \textit{mpA}(\%)       & \textbf{78.67} & 23.06 & 67.68 \\
                                &           & \# of params(KB)       & 6248.4(constant) & 33923.4(linear) & \textbf{5752.3(quadratic)} \\
                                &           & Setup time(s) & 437.20 & 85.14 & \textbf{2.50} \\\cline{2-6}
                                & \multirow{3}{*}{Large shots(50 shots)} & \textit{mpA}(\%)       & \textbf{84.20} & 66.94 & 70.39 \\
                                &                           & \# of params(KB)       & \textbf{6248.4(constant)} & 33923.4(linear) & 57523.15(quadratic) \\
                                &                           & Setup time(s) & 1973.08 & 909.22 & \textbf{25.05} \\
        \hline
    \end{tabular}
    }
    \caption{Summary for comparison of performances between gradient-based methods(\textit{TfS, LwF}) and attention-based(\textit{Att-based}) method under various settings. The numbers in the table are the average of the measurements in the range.}
    \label{table:qualitative_comparison}
\end{table*}

\begin{table*}[!t] 
    \begin{center}
        \resizebox{.9\textwidth}{!}{
        \begin{tabular}{|c|cccccc|}
            %\hline
             %\textit{snA}(\%) & \multicolumn{5}{|c|}{gradient-based model} & non-gradient-based model\\
            \hline
            % Time window(s) & Bauml et al. 2013\cite{Bauml:2013:SLC:2514950.2516130} & Tapaswi et al. 2012\cite{Tapaswi:2012} &Hu et al. 2015\cite{hu2015deep} & Ren et al. 2016\cite{ren2016look} & \multicolumn{1}{c|}{Liu et al. 2019\cite{LIU2019557}} & \textit{Att-based}(ours)\\
            Time window(s) & Bauml et al. 2013 & Tapaswi et al. 2012 & Hu et al. 2015 & Ren et al. 2016 & \multicolumn{1}{c|}{Liu et al. 2019} & \textit{Att-based}(ours)\\
            \hline
            0.5s & - & - & 74.93 & 86.59 & \textbf{87.73} & 84.34 \\
            1s & - & - & 77.24 & 89.00 & - & \textbf{92.50} \\
            1.5s & - & - & 79.35 & \textbf{90.45} & - & 88.89 \\
            2s & - & - & 82.12 & 90.84 & - & \textbf{92.45} \\
            2.5s & 77.81 & 80.80 & 82.81 & \textbf{91.17} & - & 89.89 \\
            3s & - & - & 83.42 & 91.38 & - & \textbf{93.12} \\
            
            \hline
        \end{tabular}
        }
    \end{center}
    \caption{Speaker naming accuracy(\textit{snA(\%)}) comparison between attention-based model and existing speaker naming models on real video of \textit{BBT S01E03}.}
    \label{table:sna}
\end{table*}

\subsection{Comparative Analysis among Speaker Naming Methods under Various Settings} \label{experiment:evaluation}
% difference btw previous studies and our experiments
Previous studies\cite{hu2015deep,ren2016look,LIU2019557} have evaluated the accuracy of their methods in a refined setting, which has purified voice and appears a small, fixed number of characters(5-6 IDs) in the scene. Also, they use sufficient pair embeddings per each character for training model. In this experiment, we compare our speaker naming model with existing gradient-based methods in detail under more various environments unlike previous work. By considering the advent of new characters in the story, we can precisely evaluate the performance of speaker naming methods in a more realistic situation with \textit{VoxCeleb2} dataset.
\subsubsection{Evaluation Metric}
Speaker naming is to find the matched-pairs of face and voice embeddings and predict its identity. To compare how well the speaker naming method can identify the ID of matched-pair, we defined matching pair accuracy(\textit{mpA}) as follows:
\begin{equation} \label{eqn:mpa}
    mpA = \frac{N_{id_{pred} == id_{gt}}} {N_{total}} \times 100\%
\end{equation}
% Exceptionally, we used $mpA$ for measuring identification accuracy of uni-modal embeddings to compare with multimodal embeddings in \Cref{experiment:evaluation} of the third experiment. 
The second metric is the number of parameters of speaker naming model loaded in memory. If the model consists of a neural network, the weights and biases belonged to these parameters. In the case of attention-based model, pair embeddings were counted as parameters. We convert these parameters into kilobytes(KB) and compare them.

The third metric is the setup time of model. In the case of neural network, we calculated setup time by adding data loading in memory, calculating the gradient, and updating it to the weights. In the case of an attention-based model, we measure the setup time by adding the loading time of prior knowledge embeddings and the calculation time of attention module to derive prediction results. 
\subsubsection{Experimental Setup}
% Explanation of settings in detail(FIXIT)
We conducted the experiment adjusting two main variables in situations: the number of target IDs for prediction, and the number of shots(pair embeddings of face-voice) for prior knowledge per each target ID. 
% In the case of small predicted characters, the new characters are added by 5 IDs incrementally, started from 5 to 50. In the case of large predicted characters, the new characters are added by 50 IDs incrementally, started from 50 to 500.
About the number of target IDs, we separated the situation into two parts: the number of target IDs is small or large. In each case, the number of target IDs started to be set from 5 to 50 with five increments, and from 50 to 500 with fifty increments, respectively. We also adjust the number of shots per each character set to 5(small), 50(large) shots in both situations to consider the effects of the number of labeled training data to performances of speaker naming methods.

% Explanations for metrics
% There are three metrics to be evaluated in this experiment: accuracy(\textit{mpA}), the number of parameters (converted to kilobytes) loaded in memory, and the setup time. 
%VoxCeleb2 is used
% \textit{VoxCeleb2} dataset was used for this experiment since it consists of real-world data and contains lots of individual IDs, which fits with identifying a particular ID under our assumed incremental settings.

% Introduction of baselines
As the baseline methods, we selected two representative gradient-based methods to compare with our \textit{Attention-based(Att-based)} method. The first method is \textit{Training from Scratch(TfS)}, which trains the neural network with both original and new data. Most of the deep neural networks normally use \textit{TfS} in training phase.
% Most of the deep neural networks use this method.
The second one is \textit{Learning without Forgetting(LwF)}\cite{li2017learning}, which generates the new branch on top of the network and trains with only new data.
% hyperparameter of experiments
We followed the same neural network structure with one of the previous work\cite{hu2015deep} on both methods for fair comparison. The maximum training epoch is 500, which is sufficient to converge the loss function. If the network reached the optimal cost before the maximum epoch while training, we took the accuracy and the setup time at the moment the optimal cost was derived. Transfer learning was applied in every stage of all gradient-based methods when the number of IDs was increased.

\subsubsection{Results}
%표 결과에 대한 해석
As shown in Figure \ref{fig:comp_analysis} and Table \ref{table:qualitative_comparison}, we conducted both quantitative and qualitative analysis based on the experimental results. Most notably, our method(\textit{Att-based}) significantly reduced the setup time of model compared to other gradient-based methods about tens to hundreds of times regardless of conditions. 
% It might be appropriate for the situations which occurs the addition of new character frequently, or appears various facial poses and illumination of certain character's feature embeddings which are hard to recognize without update.
The \textit{mpA} was high in order of \textit{TfS, Att-based, LwF}. However, when the number of target IDs was 450 with large shots, the \textit{LwF} gradually surpassed \textit{Att-based} as shown in the "Large IDs-$mpA$" graph in Figure \ref{fig:comp_analysis}. Generally, gradient-based methods showed a big difference in \textit{mpA} depending on the number of shots. In contrast, \textit{Att-based} worked well in both situations and had less effect in terms of the number of shots. 
\textit{Att-based} utilized small number of parameters when the number of target IDs or its shots are small. However, as the number of target IDs is increased with large(50) shots, \textit{Att-based} showed memory inefficiency because the number of parameters increased quadratically with (the number of IDs $\times$ the number of shots per ID). In contrast, \textit{TfS} occupies constant number of parameters; it is only related to the structure of the neural network. \textit{LwF} is proportional to the number of times that ID is added. Because \textit{LwF} has a multi-branch structure, the new branch is generated when the new target character comes in. 
% \textit{Att-based} scales quadratically with (the number of IDs $\times$ the number of shots per ID). Thus, \textit{Att-based} needs more parameters than other  methods as the amount of training information grows up.  

To sum up, \textit{Att-based} is the most appropriate method when new people appear frequently, and the shots per each character is not sufficient. Also, \textit{Att-based} works effectively where immediate update for hard-to-recognize data such as various facial poses is needed. Overall, \textit{TfS} is the best suited for situations where the new people are not frequently updated and high accuracy is required. \textit{LwF} locates in the middle of other two methods, because it shows faster setup of model than \textit{TfS}, but compromising its \textit{mpA} and memory usage.
% shot의 수에 따라 달라지는 그림 추가?

\subsection{Speaker Naming Accuracy for Real Video}
In this experiment, we applied our model in real video to compare the accuracy with previous gradient-based speaker naming models.

\subsubsection{Evaluation Metric}
Speaker naming accuracy(\textit{snA}) had used broadly in multiple speaker naming related papers before and was formulated in \cite{LIU2019557}, which is from one of our baseline. We used this metric to measure the performance of real video inference for comparing our model with a well-known speaker naming baselines. They define \textit{snA} as follows:
\begin{equation} \label{eqn:sna}
    snA = \frac{N_{[p^{sn} == s^{tr}]}} {N_{s^{tr}}} \times 100\%
\end{equation}
where $p^{sn}$ and $s^{tr}$ denote the labels of predicted samples and ground truth, respectively. $N_{[p^{sn} == s^{tr}]}$ is the number of correctly predicted time windows and $N_{s^{tr}}$ is the total number of time windows.

\subsubsection{Experimental Setup}
For evaluation, we followed the same settings of previous works\cite{Tapaswi:2012,Bauml:2013:SLC:2514950.2516130,hu2015deep,ren2016look} on speaker naming experiment. The four-minute-long \textit{BBT S01E03} video clip was used for evaluation dataset.
In real situation, there occurs many non-matched-pairs of face-voice embeddings per period. Unlike previous controlled settings, we put 30 shots of matched and non-matched-pairs at a ratio of 1 to 4 in prior knowledge embeddings because end-to-end inference detects not only an active speaker but also distractors.
We tested our model to video with the time window of multiple periods of 0.5s. Time windows more than 0.5s were also tested for comparing existing methods to clarify the result. If the time window is more than 0.5s, the prediction of the model is determined by the majority vote of multiple 0.5s-sized windows as previous work\cite{hu2015deep} did. 

\subsubsection{Results}
As shown in Table \ref{table:sna}, \textit{Att-based} showed comparable \textit{snA} as other gradient-based speaker naming models in most cases. In certain circumstances, such as the size of the time window is 1s, 2s, and 3s, our model even outperformed the other state-of-the-art models.  

\section{Conclusion and Future Work} \label{conclusion}
In this paper, we presented an attention-based speaker naming method for online adaptation in non-fixed scenarios. The key idea is to predict the ID of the matched-pair based on attention mechanism, which considers the correlation between all pairs of prior knowledge embeddings and extracted target embeddings.
%Then mention what is the main contribution again
Our proposed approach significantly reduced the model setup time by keeping comparable accuracy to existing state-of-the-art models, as demonstrated in our experiments. Also, the model can be updated online by only changing information on the attention module.

Our further research aims to solve the current limitations and improve the method well-applied to more generalized situations. Now, our current method was using only two modalities and showed low accuracy when the number of target IDs for identification is large. Also, it can occur memory inefficiency if the number of IDs and the number of shots per ID are increased. If we properly combine the advantages of the gradient-based methods with our method, the integrated method will be one of the solutions to cover more various situations adequately in the future.

%ACKNOWLEDGMENTS are optional
\section{Acknowledgments}
This work was supported by the New Industry Promotion Program(1415158216, Development of Front/Side Camera Sensor for Autonomous Vehicle) funded by the Ministry of Trade, Industry \& Energy(MOTIE, Korea).

\bibliographystyle{aaai} \bibliography{references.bib}
\end{document}